\documentclass{aip-cp}

\usepackage[numbers]{natbib}
\usepackage{rotating}
\usepackage{graphicx}
\newcommand{\be}{\begin{equation}} 
\newcommand{\ee}{\end{equation}}
\newcommand{\bc}{\begin{center}}
\newcommand{\ec}{\end{center}}

\begin{document}

\title{The algebraic molecular model in $^{12}$C and its application to the $\alpha$+$^{12}$C scattering: from densities and transition densities to optical potentials and nuclear formfactors}

\author[aff1,aff2]{Andrea Vitturi\corref{cor1}}
\author[aff1,aff2]{Jesus Casal}
\author[aff1,aff2]{Lorenzo Fortunato}
\author[aff3]{Edoardo G. Lanza}

\affil[aff1]{Dipartimento di Fisica e Astronomia, I-35131 Padova, Italy}
\affil[aff2]{INFN, Sezione di Padova, I-35131 Padova, Italy}
\affil[aff3]{INFN, Sezione di Catania, I-95123, Catania, Italy}
\corresp[cor1]{Corresponding author: vitturi@pd.infn.it}

\maketitle

\begin{abstract}
The algebraic molecular model is used in $^{12}$C to construct densities and transition densities connecting low-lying states of the rotovibrational spectrum, first and foremost those belonging to the rotational bands based on the ground and the Hoyle states.  These densities are then used as basic ingredients to calculate, besides electromagnetic transition probabilities, nuclear potentials and formfactors to describe elastic and inelastic $\alpha$+$^{12}$C scattering
processes.  The calculated densities and transition densities are also compared with those obtained by directly solving the problem of three interacting alpha's within a three-body approach where continuum effects, relevant in particular for the Hoyle state, are properly taken into account. 

\end{abstract}
\section{INTRODUCTION}
The low-lying structure of $^{12}$C is still one of the most fascinating open problems in nuclear physics.  The issue of  alpha-clusterization and the nature of the first excited 0+ state (the so-called Hoyle state, that plays a fundamental role in the nucleosyntesis processes) are two highly interesting issues.  Different probes have been extensively used in these studies.  We want here to concentrate on inelastic nuclear excitation, as for example in $\alpha$+$^{12}$C scattering
processes, with particular focus on the population of the states in the Hoyle band and the information that this could provide on the radial extension of the Hoyle state.  The necessary structure inputs will be determined within the algebraic cluster model by Iachello and collaborators \cite{Bij02,Bij14,Bij95,Del17,Stel}, that conveniently and elegantly reshapes the seminal alpha cluster molecular model by Wheeler \cite{Wee}.
   
\section{DENSITIES AND TRANSITION DENSITIES}

Within the molecular model the intrinsic density associated with the ground state band of $^{12}$C can be expressed as a sum of three $\alpha$ particles placed at the vertices of a triangle, each particle being displaced of the proper amount, $\beta$.  We have therefore 
\be
\rho_{A_0}(\vec r, \{ \vec r_k\},\beta)= \sum_{k=1}^3 \rho_\alpha(\vec r - \vec r_k)
\label{dez}
\ee
with $\vec r_1=(\beta,\pi/2,0)$, $\vec r_2=(\beta,\pi/2,2\pi/3)$ and  $\vec r_3=(\beta,\pi/2,4\pi/3)$ in spherical polar coordinates $(r,\theta,\phi)$, where the co-latitude is always $\pi/2$ because we have chosen a triangle lying in the $\{xy\}$ plane with the particle labeled as $1$, lying on the positive $x-$axis.
The density of each $\alpha$ particle is taken as a gaussian function:
\be
\rho_\alpha(\vec r)= \Bigr(\frac{\alpha}{\pi}\Bigl)^{3/2} e^{-\alpha r^2}
\ee
with $\alpha=0.56(2)$ fm$^{-2}$ as in Ref. \cite{Del17}. 
This `static' ground-state density is labeled with $A_0$ because its shape is associated with the fully symmetric representation of $D_{3h}$ with $0$ quanta of excitation. It is displayed as a contour plot in the left frame in Fig.~1.
The value $\beta=1.82$ has been adopted for the radial parameter (at variance with the value $\beta=1.74(4)$ fm used in Ref. \cite{Del17}) to optimize the fit to both ground state radius and the $B(E2)$ to the first excited $2^+$ state. 
It can be expanded in spherical harmonics as
\be
\rho_{A_0}(\vec r, \beta)= \sum_{\lambda\mu} \rho_{A_0}^{\lambda,\mu} (r)~Y_{\lambda,\mu} (\theta,\varphi)
\ee
where the radial functions depend on $\lambda,\mu$. Our choice of coordinates is such that, only those multipoles $\lambda,\mu$ that are allowed by the $D_{3h}$ symmetry appear in the sum.  The lowest terms in the sum are $\lambda,\mu$=(0,0), (2,0) and (3,3). This is different from Ref. \cite{Del17} where the $z-$axis was instead chosen to pass through particle $1$ and the center of the triangle. The left frame of Fig.~2 shows the three lowest order radial functions of the expansion in spherical harmonics for $\{\lambda\mu\}=\{00,20,33\}$.  From these intrinsic densities one can obtain densities and transition densities in the laboratory frame associated with the lowest states belonging to the ground-state rotational band.  For example the intrinsic function labeled $00$ gives directly the ground state density, while the others represent the radial transition densities associated with transitions
connecting the ground state with the different member of the ground state band (black lines in Fig.~3).
\begin{figure}[h]
\includegraphics[width=0.33\textwidth,clip=]{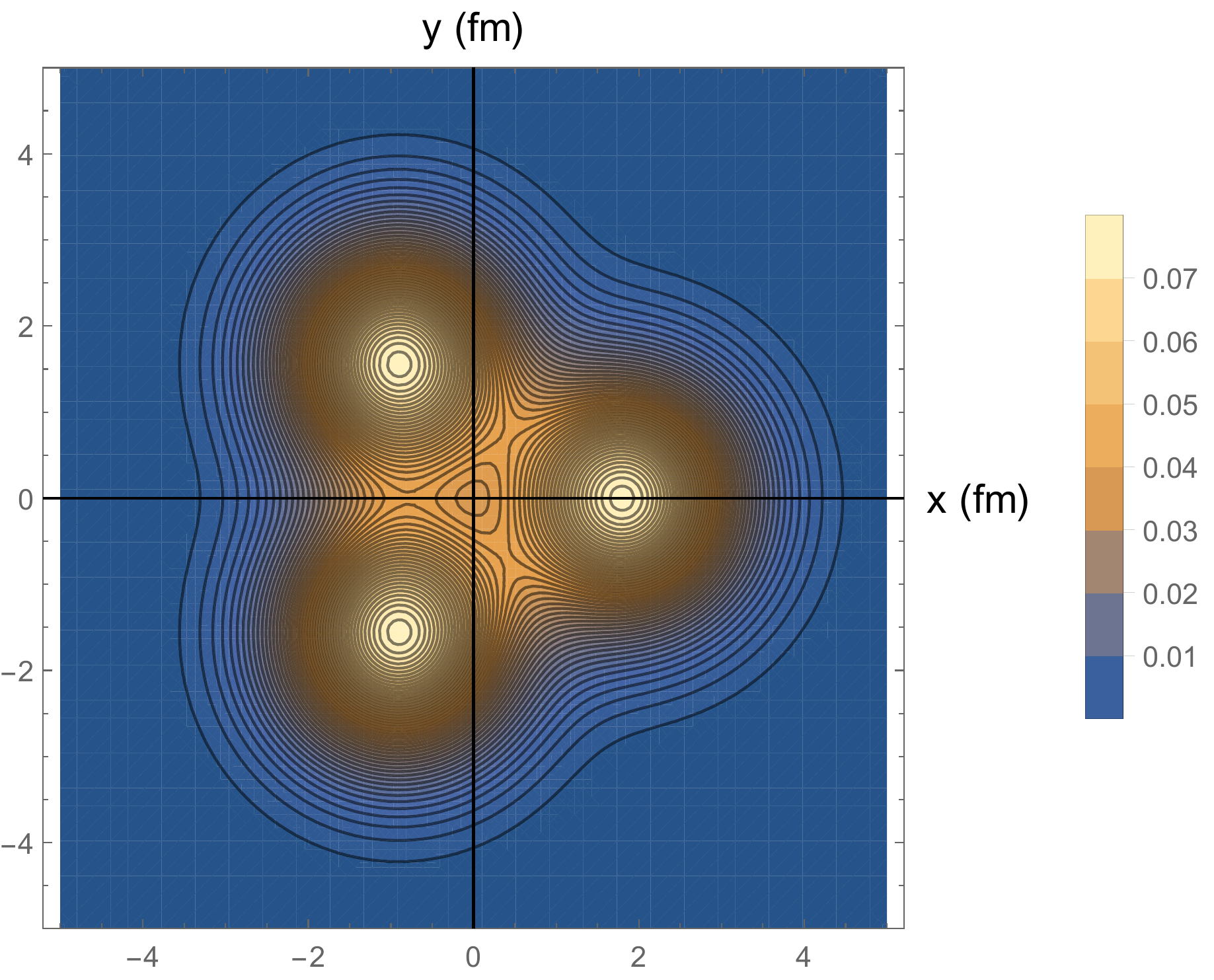}
\includegraphics[width=0.33\textwidth,clip=]{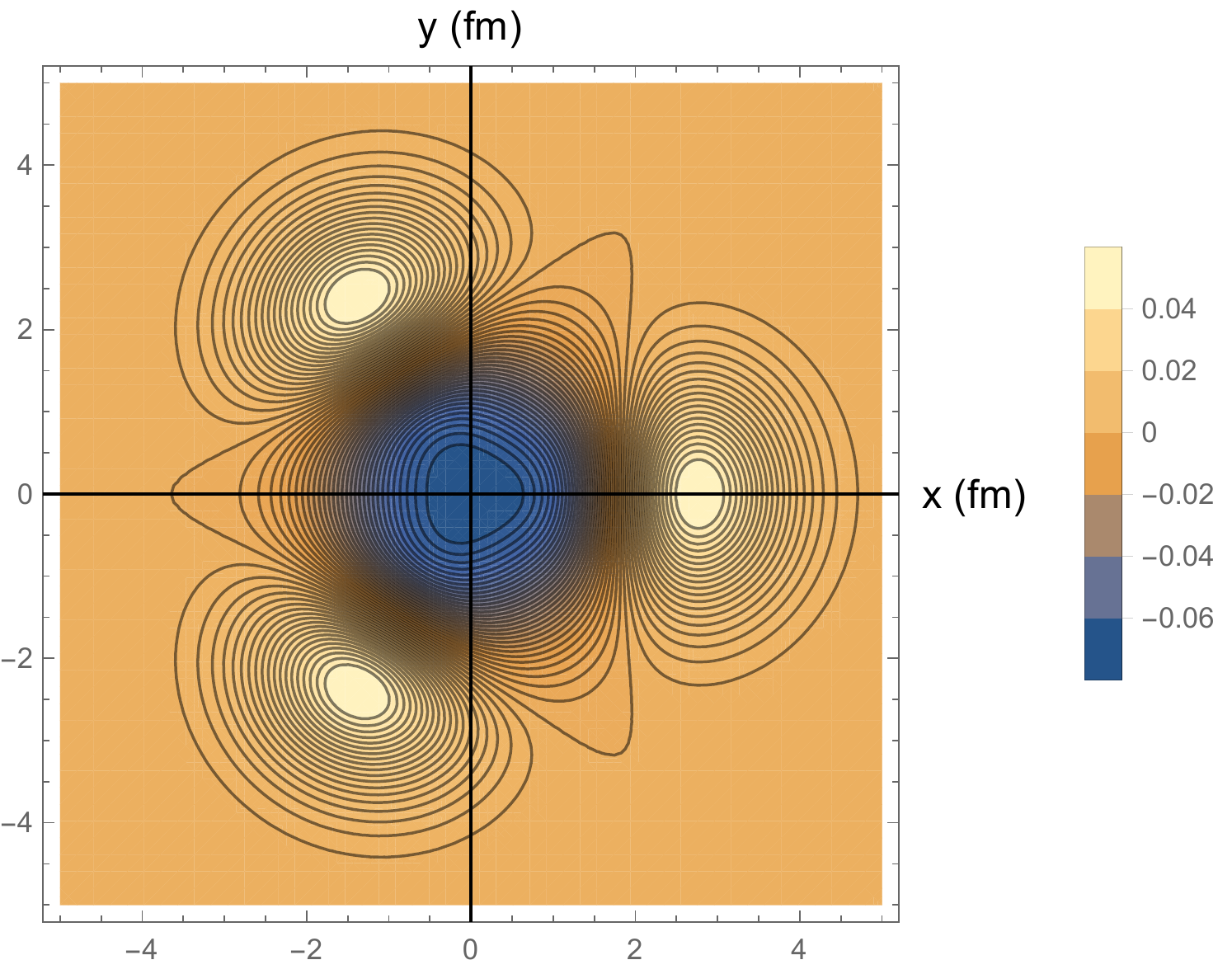}
\includegraphics[width=0.33\textwidth,clip=]{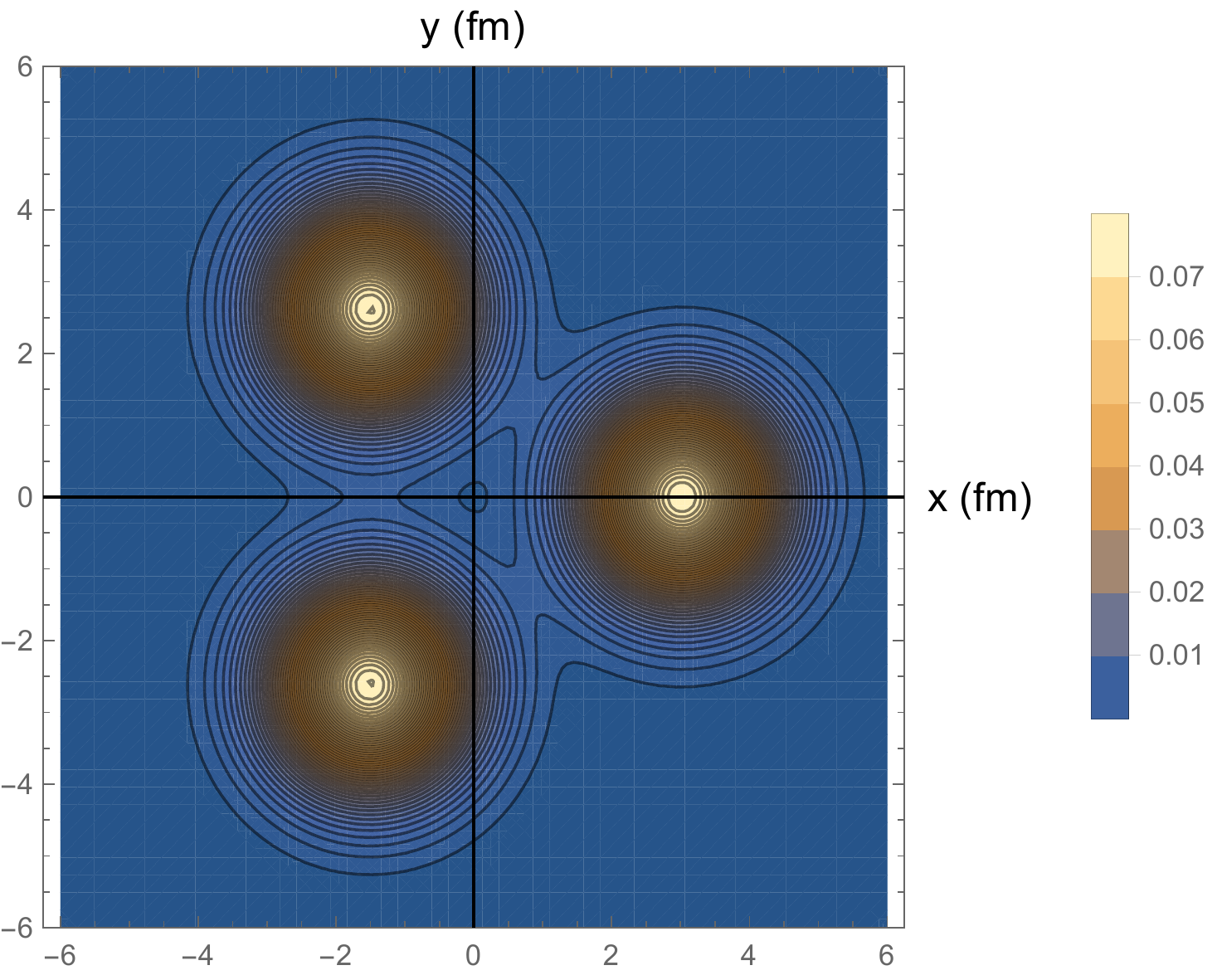} 
\caption{Left: Contour plot of the intrinsic density of the g.s. static triangular configuration. This is a cut on the $z=0$ plane. Middle: Contour plot of the intrinsic transition density connecting g.s. and Hoyle-band intrinsic states. Right: Contour plot of the intrinsic density of the Hoyle band.}
\end{figure}
\begin{figure}[h]
\includegraphics[width=0.33\textwidth,clip=]{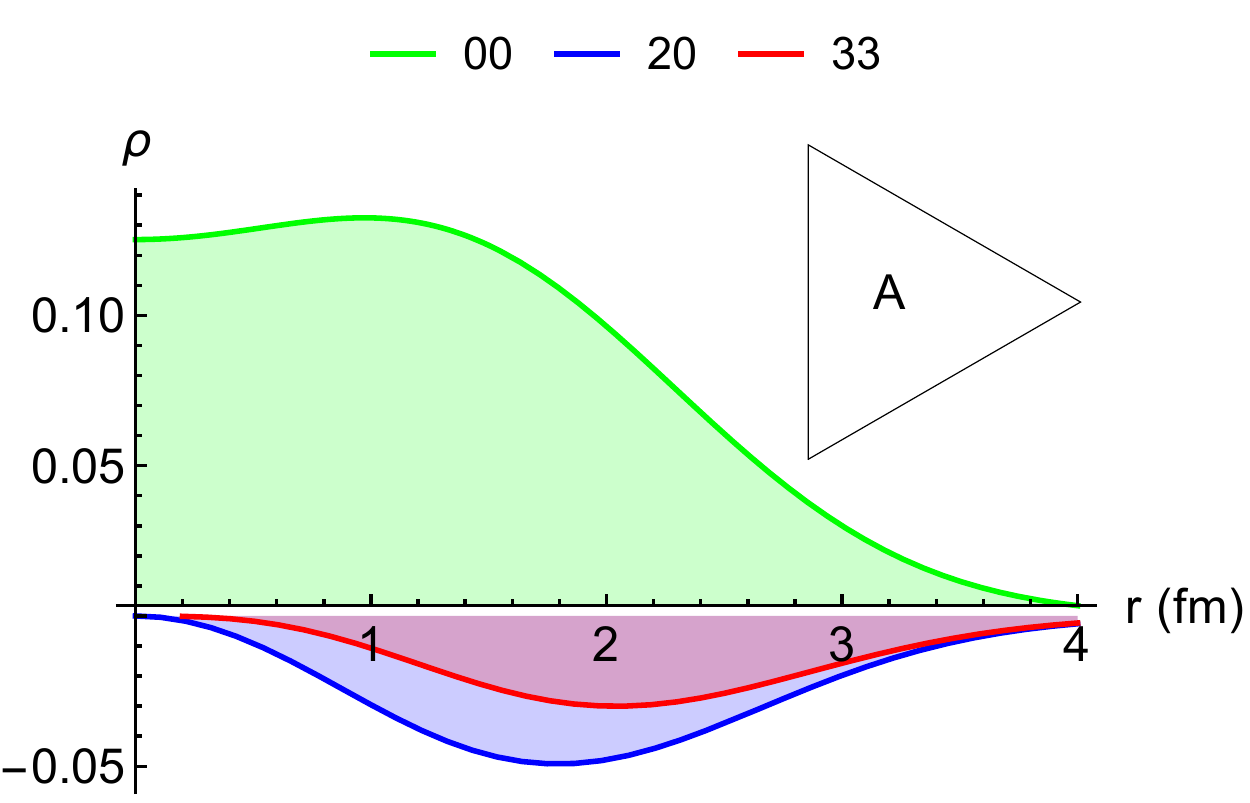}
\includegraphics[width=0.33\textwidth,clip=]{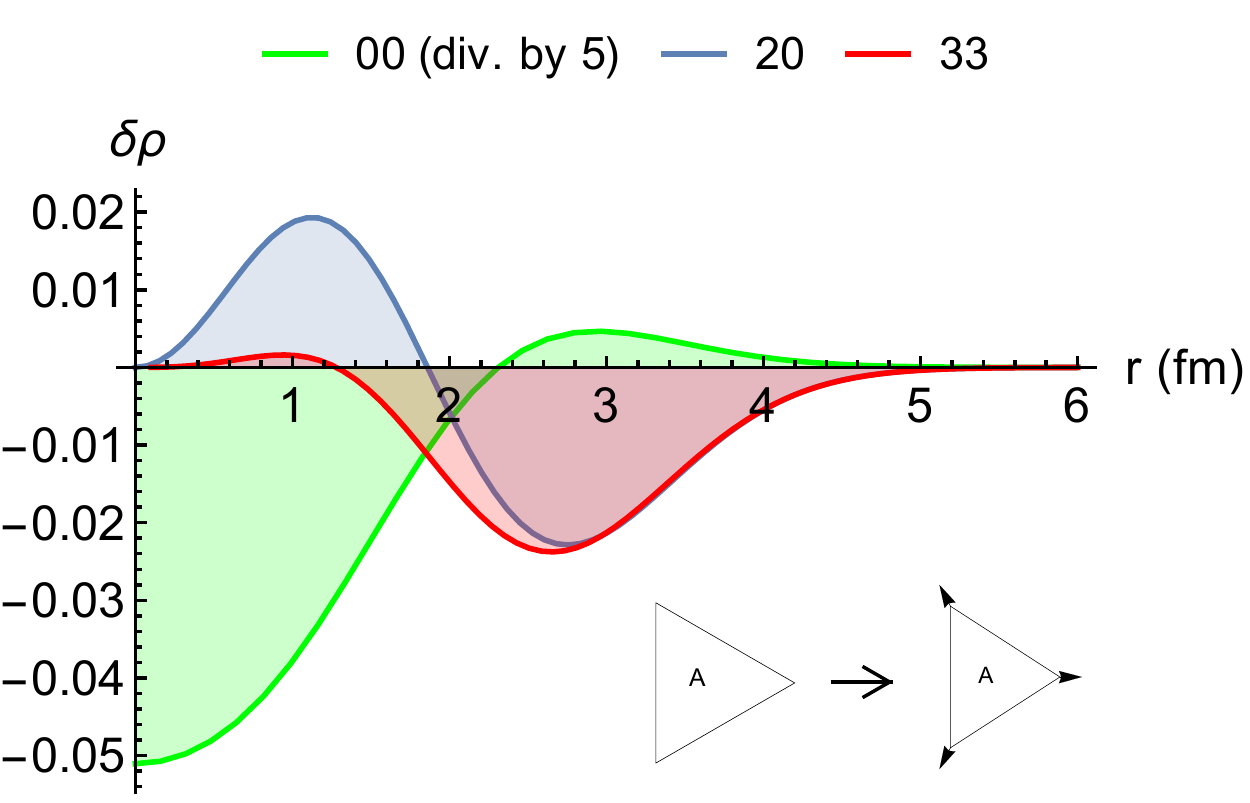}
\includegraphics[width=0.33\textwidth,clip=]{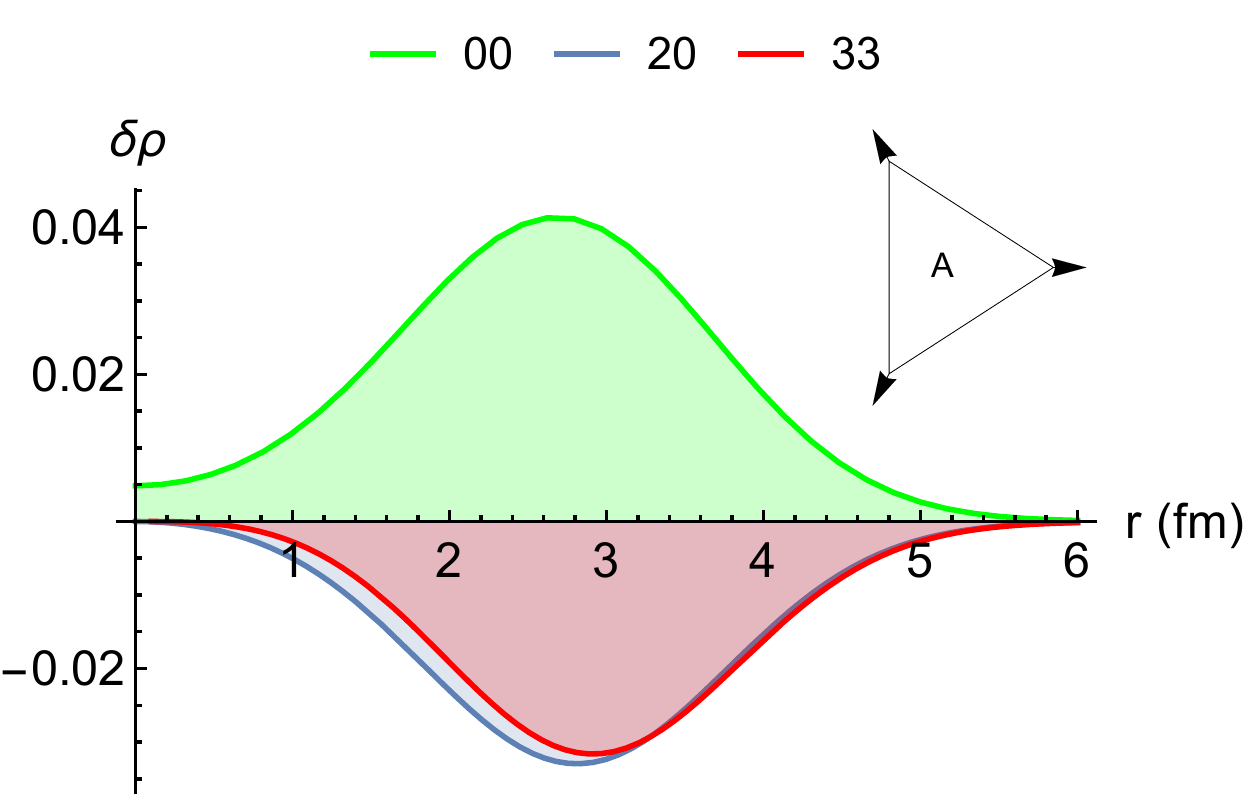} \caption{Radial components (for lower angular momenta) of the expansion in spherical harmonics of the intrinsic g.s. density (left panel), the intrinsic transition density connecting the g.s. and Hoyle-state band intrinsic states (middle panel), and the intrinsic density of the Hoyle band (right panel). }
\end{figure}
\begin{figure}[h]
\includegraphics[width=0.4\textwidth,clip=]{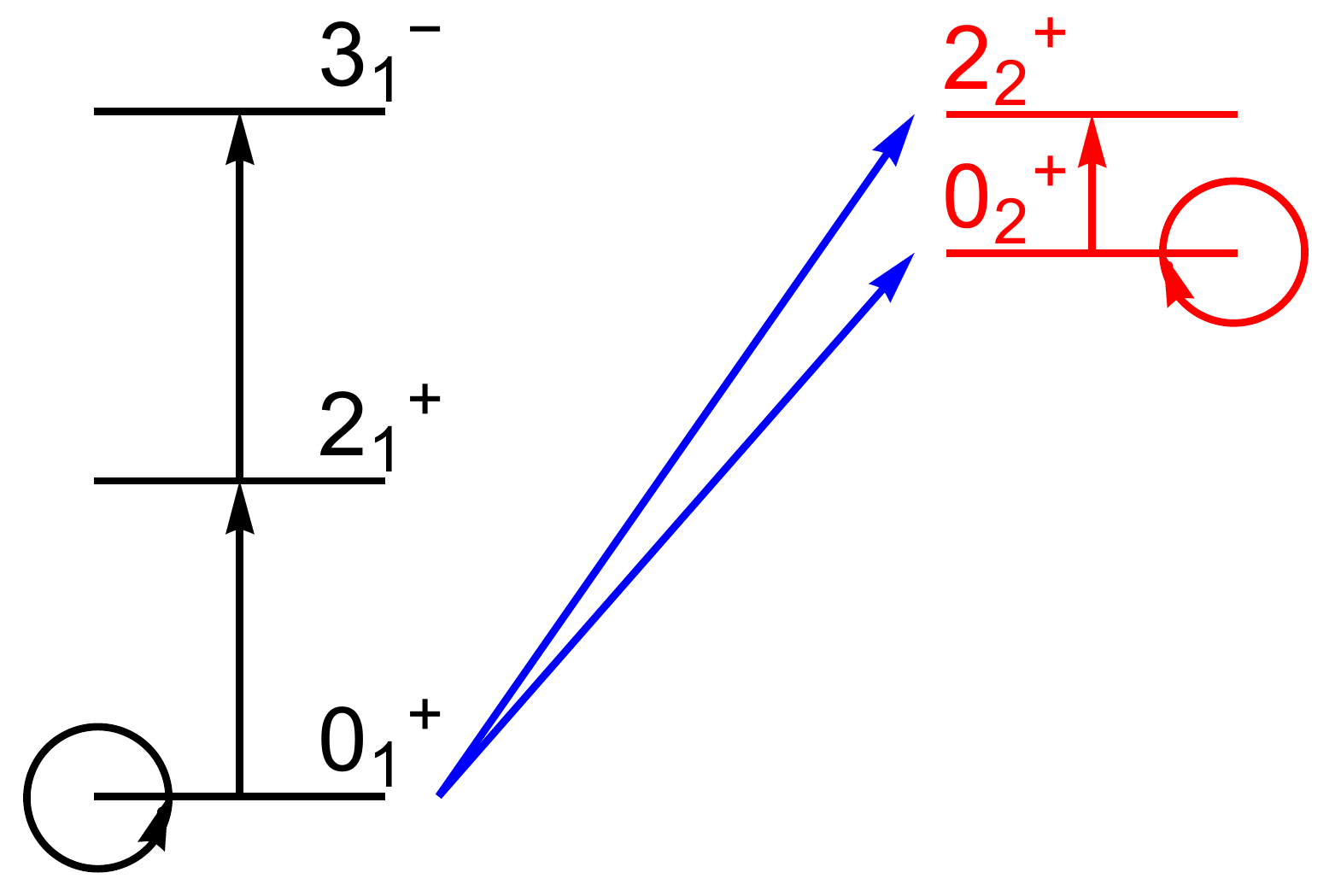}
  \caption{Schematic pictues of the lowest states and the lowest intra- and inter-band transitons}
\end{figure}
\begin{figure}[h]
\includegraphics[width=0.5\textwidth,clip=]{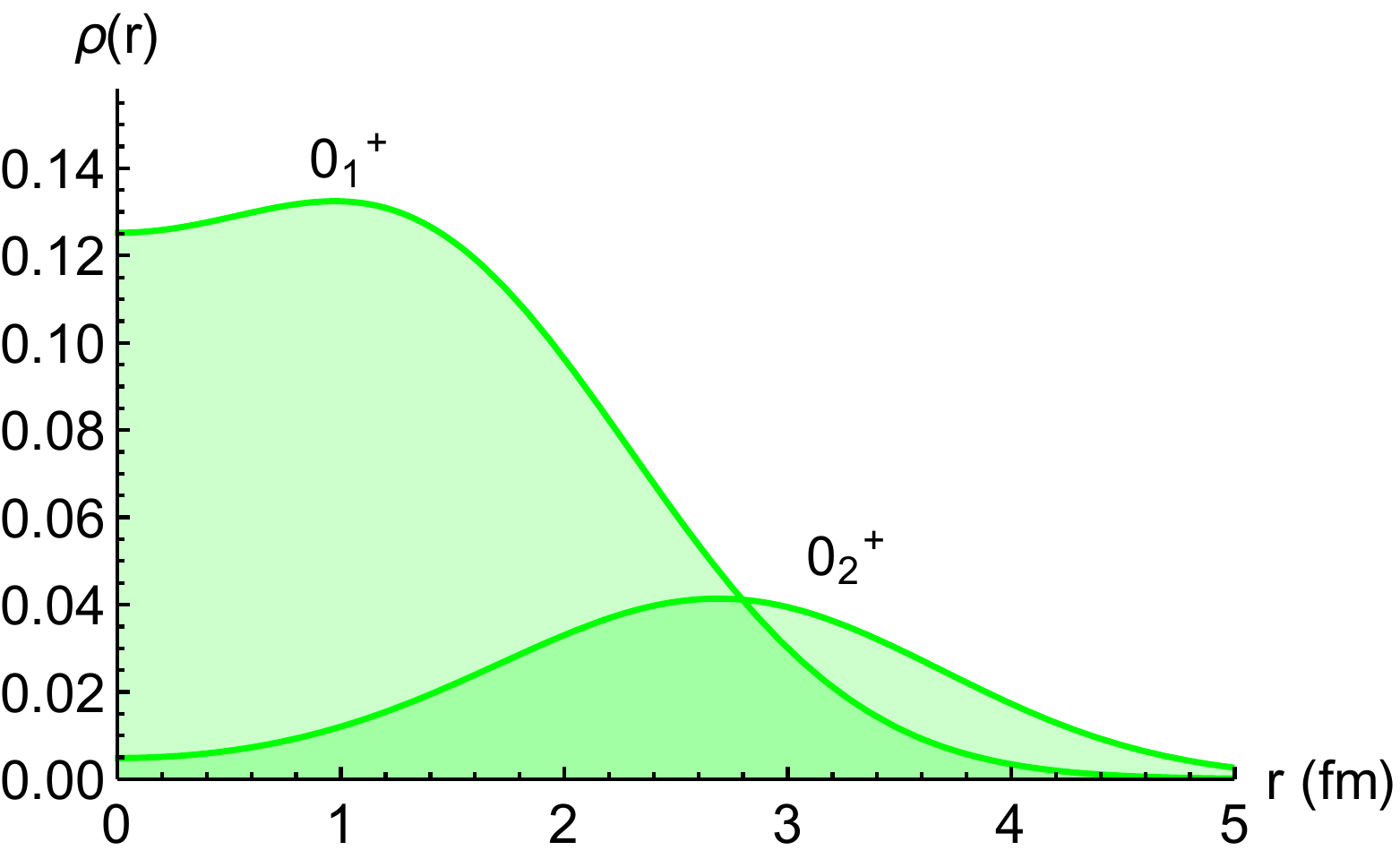}
\includegraphics[width=0.5\textwidth,clip=]{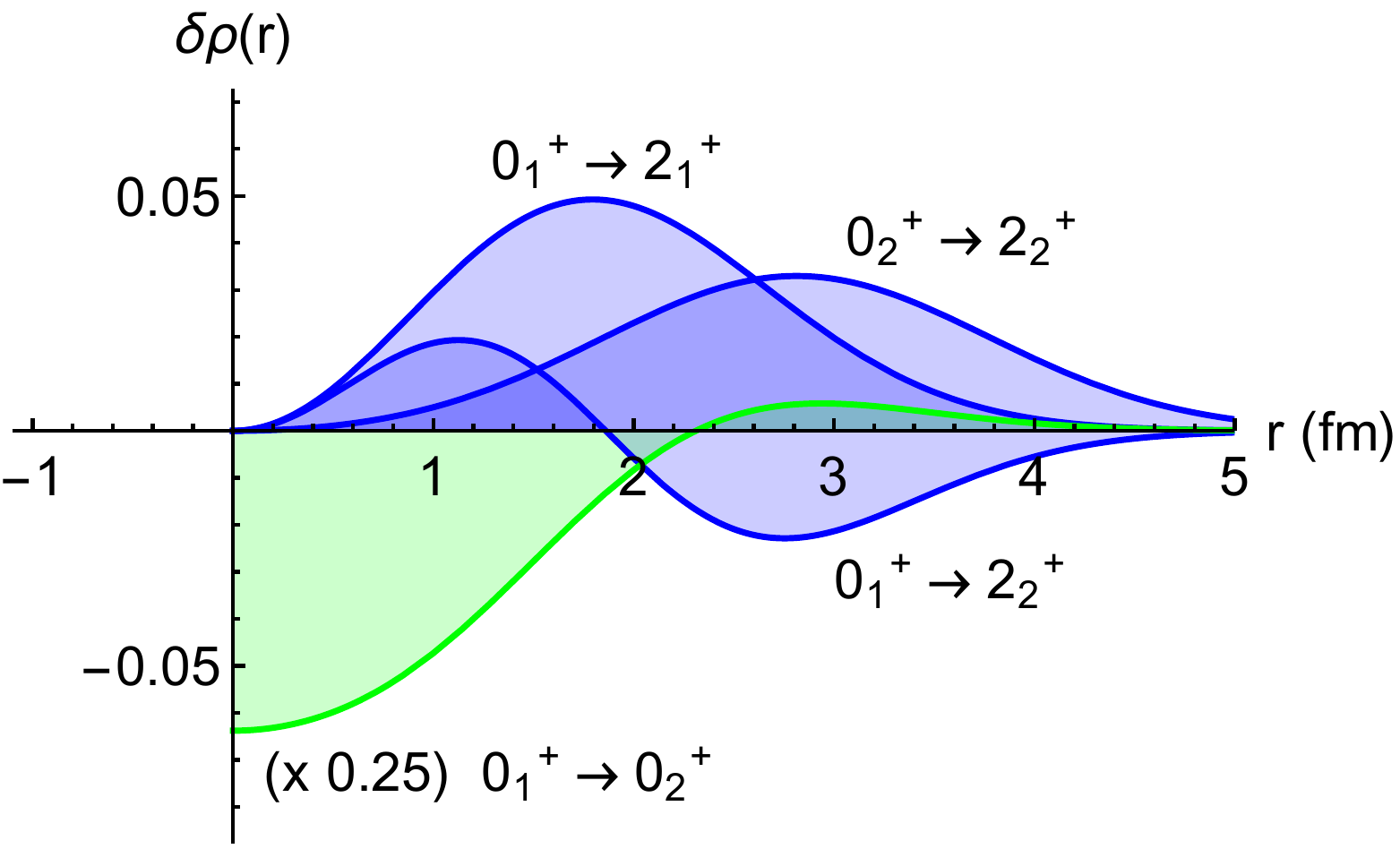}
  \caption{Left: Radial densities of the ground and Hoyle states. Right: Interband and intraband quadrupole transition densities (0$^+_1$ to 
 2$^+_1$, 0$^+_1$ to 
 2$^+_2$ and 0$^+_2$ to 
 2$^+_2$.  The monopole transition density connecting 0$^+_1$ to 
 0$^+_2$ is also given.}
\end{figure}

Within the molecular models excited bands are associated to vibrations of the three alpha particles along the directions of the vectors of normal modes of motion, that are of two types: singly-degenerate fully-symmetric, $A$, and doubly-degenerate, $E$. For example we can obtain the first symmetric vibrational band of $A-$type  (with $n=1$), corresponding to the band constructed over the ``Hoyle state'' (labeled $A_1$,  by considering small symmetric displacements  $\Delta \beta^A$ along the radial direction (breathing mode).  The corresponding intrinsic transition density connecting ground and Hoyle bands can be obtained, in leading order in $\Delta \beta^A$, in the form
\be
\delta \rho_{A_0\rightarrow A_1} (\vec r) =\chi_1\frac{d}{d\beta}\rho_{A_0}(\vec r, \beta)~.
\ee
The intrinsic transition matrix element $\chi_1$ has been set equal to $\chi_1=0.247$ fm using the value of the matrix element $M(E0)$ (as measured in Ref. \cite{Stre}).
The deformed transition density from the ground band to the Hoyle band is displayed as a contour plot in the central frame in Fig.~1.
As in the previous case the transition density can be expanded in spherical harmonics as
\be
\delta\rho_{A_0}(\vec r, \beta)= \sum_{\lambda\mu} \delta\rho_{A_0\rightarrow A_1}^{\lambda,\mu} (r) ~Y_{\lambda,\mu} (\theta,\varphi)
\ee
Due to the same symmetry properties the lowest terms in the sum are again $\lambda,\mu$=(0,0), (2,0) and (3,3) and the central frame of Fig.~2 shows the three lowest order radial functions of the expansion in spherical harmonics.  From these intrinsic transition densities one can obtain the interband transition densities in the laboratory frame connecting states in the ground-state and Hoyle-state rotational bands.  For example the intrinsic  functions give directly  the radial transition densities associated with transitions connecting the $0^+_1$ with the different members of the Hoyle- state band (blue lines in Fig.~3).

In analogy to the ground band, we can construct the intrinsic density associated to the Hoyle band.  Following the symmetry properties of this band, we obtain the intrinsic density by placing the three alpha's at the vertices of a triangle.  Due to the vibrational nature of the band the effective radial parameter of the three alpha's is larger than in the ground band.  In our calculation we have chosen a value of $\beta_{Hoyle}$=3.02 fm in order to get a density radius and a B(E2) value from $0^+_2$ to $2^+_2$ along the findings of microscopic approaches.  The intrinsic density is displayed as a contour plot in the right frame in Fig.~1.
As in the previous cases the intrinsic density can be expanded in spherical harmonics and the corresponding lowest-order multipole radial terms are shown in the right frame of Fig.~2.  Again the intrinsic function labeled $00$ gives directly the density of the band-head of the Hoyle band, while the others represent the radial transition densities associated with intraband transitions
connecting Hoyle state with the different member of its band (red lines in Fig.~3).

The different size associated to the ground and Hoyle bands can be appreciated from the comparisons shown in Fig.~4.  In the left frame we compare the radial densities of the ground and Hoyle states.  The larger radial extension of the Hoyle state is evident, as well as its predicted central depletion.  Similar effects can be seen in the quadrupole transition densities that are compared in the right frame. 
The intraband 0$^+$ to 2$^+$ transition densities have a similar shape in the two bands (surface peaked), but the one of the Hoyle band peaks at a rather larger radius.  On the other hand the interband quadrupole transition density and the monopole transition density connecting 0$^+_1$ and 0$^+_2$ have a node on the surface and this will be reflected in the fact that the transition densities in the tail (which is the most relevant region in the scattering process) do not scale as the corresponding matrix elements involving the transtion densities. 

\section{POTENTIALS AND NUCLEAR FORM FACTORS FOR $\alpha$+$^{12}$C SCATTERING}

All the structure information described above will then be used to describe the inelastic excitation of the low-lying spectrum in $^{12}$C.
As a test case we will consider the $\alpha$+$^{12}$C scattering process.  The complicated structure calls for the description of the reaction within a full coupled-channel scheme and this implies, as a first step, the need for the construction of diagonal potentials and coupling formfactors. 
The real part of the optical potential is constructed with the double folding procedure\cite{sat,sat1} in the form
\be
V_{N}(r)~=~\int\int \rho_{_P}(r_1) ~\rho_{_T}(r_2)~ v_0(r_{12})~ d{\bf r_1} d{\bf r_2}
\ee
in terms of the projectile ($\rho_{_P}$) and target ($\rho_{_T}$) densities and the effective nucleon-nucleon interaction $v_0$.  Given the "isoscalar" nature of the projectile $\alpha$ (N=Z) only the isoscalar part of the nucleon-nucleon interaction gives non vanishing contributions. For $v_0$ we use the density independent M3Y nucleon-nucleon interaction, Reid type\cite{m3y}, whose explicit expressions can be found in ref.\cite{sat}. It has been successfully used for the description of many elastic and inelastic reactions. The implementation of a density dependence shows that the obtained potentials have small difference in the interior of the nucleus and are very similar in the region of the nuclear surface\cite{khoa}. 
The two nuclear potentials associated with the ground and the Hoyle band states are shown in the left panel of Fig.~5. They have been constructed by using the densities plotted in the left panel of Fig.~4 and the density of the $\alpha$ particle of Eq.~2.  Following the different ranges of the densities the potential associated with the Hoyle state has a longer tail with respect to the one of the ground state.
\begin{figure}[h]
\includegraphics[width=0.5\textwidth,clip=]{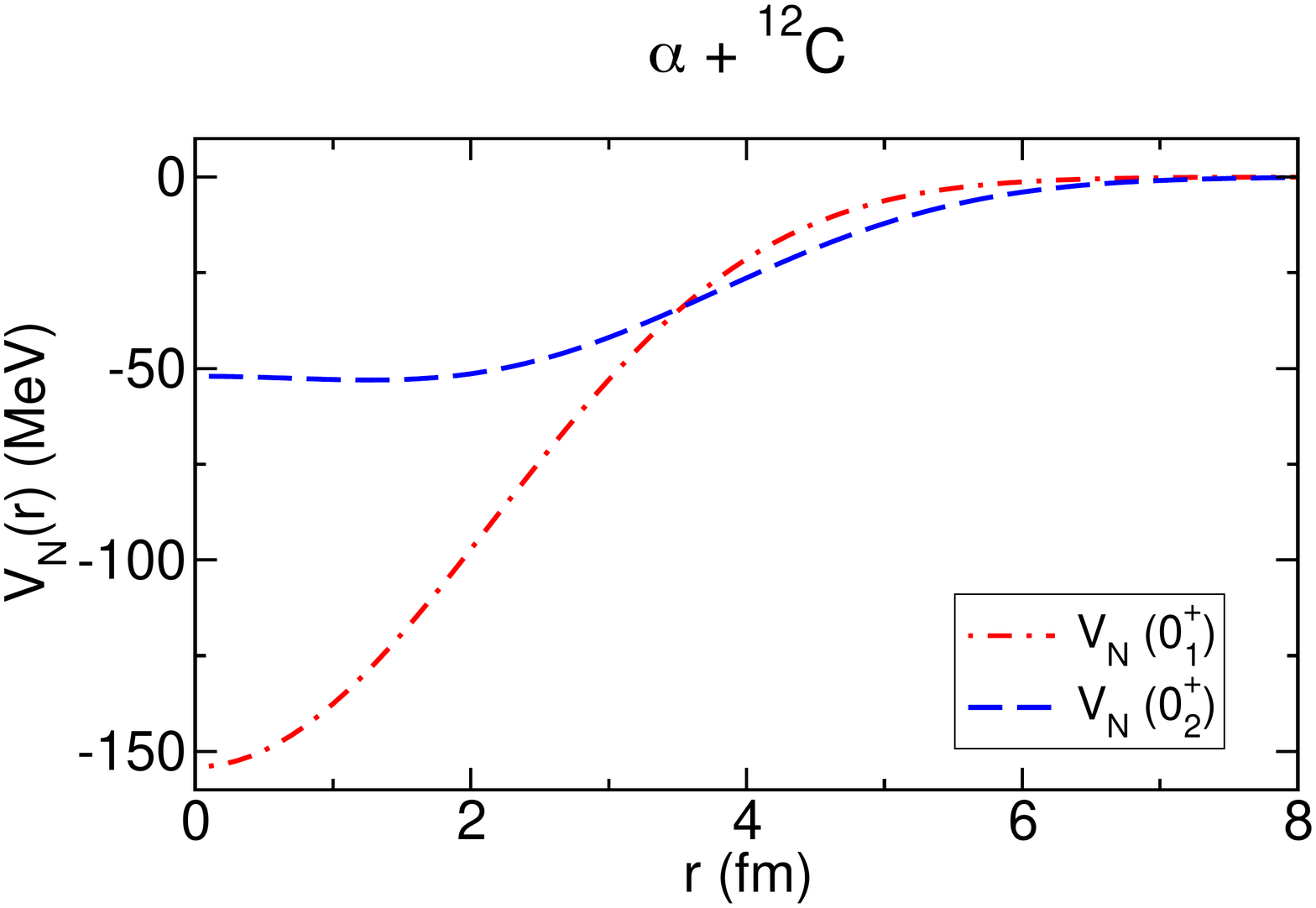}
\includegraphics[width=0.5\textwidth,clip=]{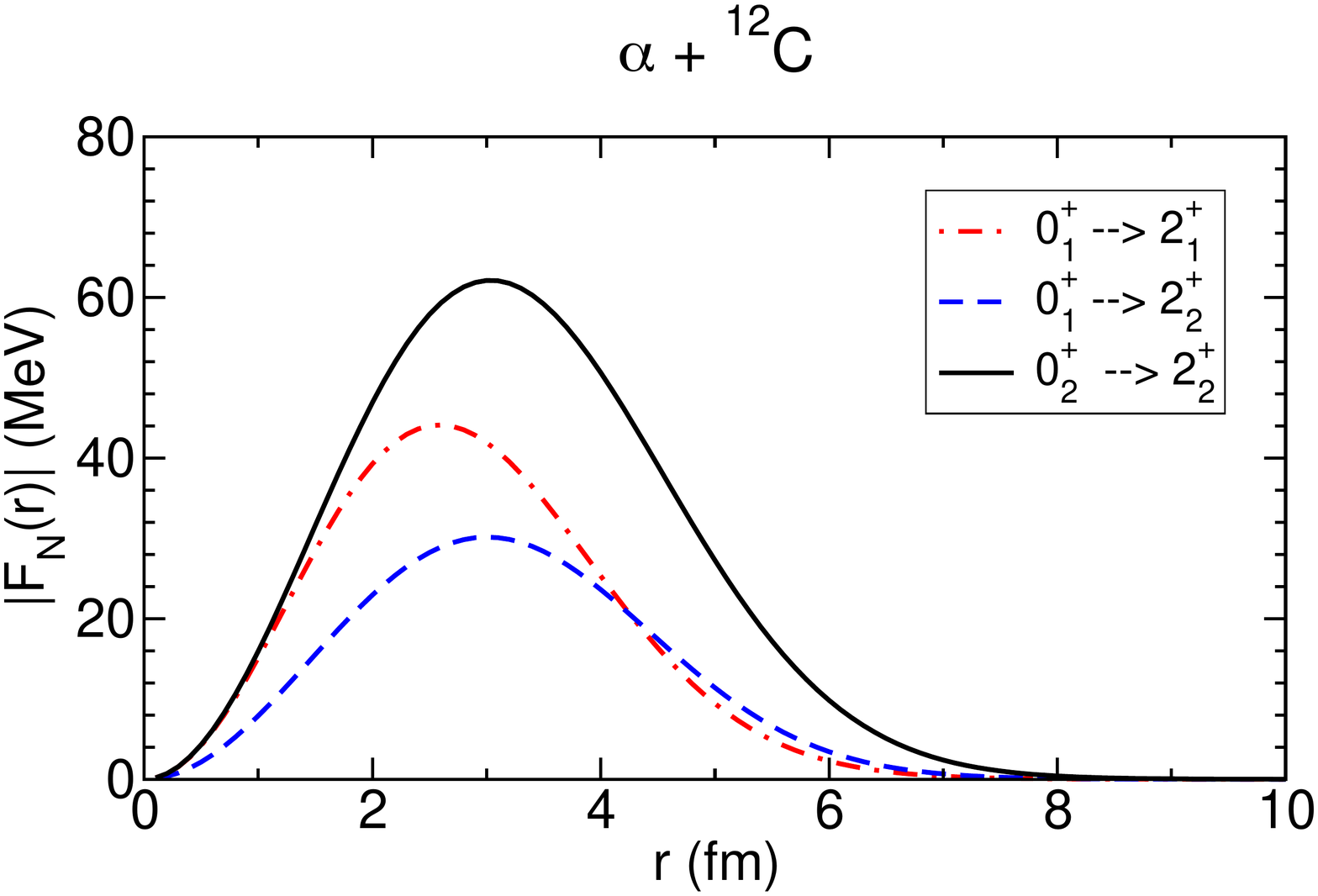}
  \caption{Left panel: Double folded nuclear potentials for the system $\alpha$ +$^{12}$C for the ground (red dot-dashed line) and the Hoyle (blue dashed line) band states. Right panel: Absolute value of radial form factor for the system $\alpha$ +$^{12}$C for the three excitation processes shown in the legend. The quadrupole states are the ones built on top of the ground state (2$^+_1$) and on top of the 0$^+_2$ Hoyle state (2$^+_2$)}
\end{figure}

The transition densities are the basic ingredients to construct the
nuclear formfactors describing nuclear excitation processes.  These
formfactors can again be obtained by double folding\cite{sat,sat1} the
transition densities of the target with the ground state density of the projectile and with the same effective
nucleon nucleon interaction used for the potential. We include only the isoscalar terms because also in this case having N=Z implies that the isovector term is zero. Proceeding as in the case of the potential then the following expression for the form factors are obtained 
\be
F_{N}=\int\int \rho_{_P}(r_1)~
v_0(r_{12})~ \delta\rho_{_T}(r_2)~ d{\bf r_1} d{\bf r_2}~
\ee

The excitation processes of interest are the ones related to the first 2$^+$ state in both ground and Hoyle band and the comparison among the intra- and inter-band transitions. In the right panel of Fig.~5 the radial form factors for the three most interesting transitions are shown. As one can infer from the legend, these are the ones from the ground state to the 2$^+$'s built on the ground state (2$^+_1$) or on the Hoyle state (2$^+_2$) together with the one going from the Hoyle state (0$^+_2$) to the (2$^+_2$). We note that the form factor for the transition from 0$^+_2$ to 2$^+_2$ has a larger radial extension  that the other two transitions taken into consideration. As a consequence the angular distribution for this transition may extend on a reduced range compared to the other ones and therefore might give a hint on the radial extension of the 2$^+_2$ state\cite{ito}.  In addition the strong inband coupling could give rise to a significant interference between the direct population of the  2$^+_2$ state and the two-step process via the 0$^+_2$ state.  A detailed study of the features of this interference, as a function of the scattering angle and of the bombarding energy, should give information on the different radial size of the ground and Hoyle bands.  Results of the coupled channel calculations using the previous formfactors for the $\alpha$+$^{12}$C scattering will be given in a forthcoming paper.

\section{A comparison with the model of three interacting alpha's}
\begin{figure}[h]
\includegraphics[width=1\textwidth]{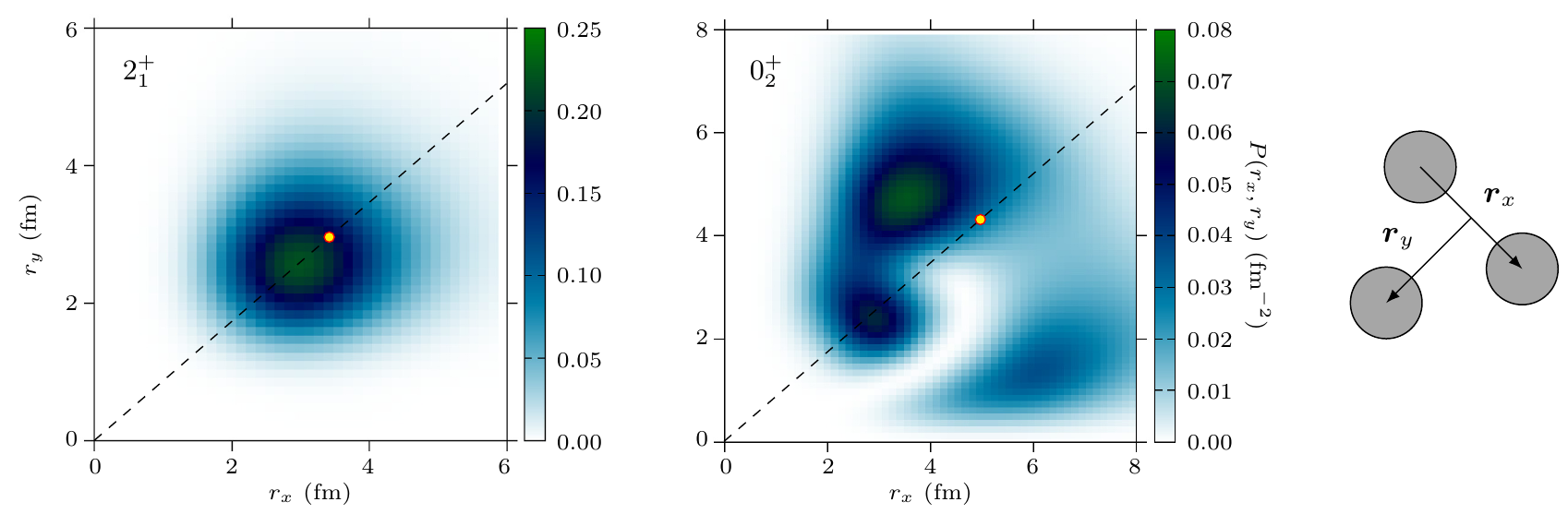}
  \caption{Probabilities of the 2$^+_1$ and 0$^+_2$ states as a function of Jacobi coordinates $\{r_x,r_y\}$. Dashed lines represent the equilateral ratio $r_y= r_x \sqrt{3}/2$ and the yellow dots correspond to the r.m.s.~values.}
  \label{fig:3bprob}
\end{figure}

Microscopic theories sustain three-alpha cluster configurations for the lowest states in $^{12}$C~\cite{Chernykh07}. Apart from its ground state, all known states in $^{12}$C lay close to the $\alpha+\alpha+\alpha$ threshold or above, which makes strict three-body models a suitable representation of the system~\cite{Nguyen13}. Thus, in order to establish a robust foundation to algebraic approaches, we are studying the symmetry of $^{12}$C in a model of three interacting alpha's. For this purpose, we solve the problem of three identical $S=0$ bosons in hyperspherical coordinates using a pseudostate method. This approach provides the bound states of the system, but also a discrete representation of the continuum (e.g.~Ref.~\cite{JCasal16} for details). In this work, as in Refs.~\cite{Nguyen13,Ishikawa14}, we parametrize the alpha-alpha interaction as an Ali-Bodmer potential~\cite{AliBodmer}. The Hamiltonian is then diagonalized in a large basis of transformed harmonic oscillator functions. Here, we have focused on the $2_1^+$ excited bound state and the $0^+_2$ Hoyle state. The latter, although unbound, can be approximated as a single pseudostate given its extremely narrow width. The corresponding probability distributions, in Jacobi coordinates, are shown in Fig.~\ref{fig:3bprob}. By studying the angle between Jacobi coordinates, it can be shown that these probabilities yield configurations of isosceles triangles~(e.g.~Ref.~\cite{Ishikawa14}). This gives a a microscopic basis to algebraic models. For the $2^+_1$ bound state, we find a single maximum with $r_y\simeq r_x \sqrt{3}/2$, which corresponds to an equilateral triangle. For the $0^+_2$ Hoyle state, on the contrary, three different structures appear in the probability plot. The dominant one corresponds to an $\alpha$ particle far from the other two, a configuration sometimes called prolate triangle~\cite{Nguyen13}. However, it is worth noting that the mean value of $r_x$ and $r_y$ for the Hoyle state satisfies also the ratio corresponding to an equilateral triangle, which indicates that the overall triangular symmetry is somewhat valid. The computation of transition amplitudes and related observables in a three-body model is being carried out and will be presented elsewhere.

\section{ACKNOWLEDGMENTS}
It is a pleasure to dedicate this work to Franco Iachello who has been for all these years an infinite source of inspiration and suggestions.

\vskip1truecm

\bibliographystyle{aipnum-cp}%
\renewcommand{\section}[2]{}

\end{document}